\documentclass[prl,aps,twocolumn]{revtex4}

\usepackage{amsfonts}
\usepackage{amssymb}
\usepackage{graphicx}
\usepackage{bbold}
\usepackage{textcomp}
\usepackage{dcolumn}    
\usepackage{bm}         
\usepackage{hyperref}

\begin{document}

\title{Detection and control of individual nuclear spins using a weakly coupled electron spin}

\author{T.~H.~Taminiau$^1$, J. J. T. Wagenaar$^1$, T. van der Sar$^1$,\\ F. Jelezko$^2$, V. V. Dobrovitski$^3$, and R. Hanson$^1$}
\affiliation{$^1$Kavli Institute of Nanoscience, Delft University
of Technology, PO Box 5046, 2600 GA Delft, The Netherlands.\\
$^2$Institut f\"{u}r Quantenoptik, Universit\"{a}t Ulm, 89081 Ulm,
Germany.\\ $^3$Ames Laboratory and Iowa State University, Ames,
Iowa 50011, USA.}

\date{\today}

\begin{abstract}
We experimentally isolate, characterize and coherently control up
to six individual nuclear spins that are weakly coupled to an
electron spin in diamond. Our method employs multi-pulse sequences
on the electron spin that resonantly amplify the interaction with
a selected nuclear spin and at the same time dynamically suppress
decoherence caused by the rest of the spin bath. We are able to
address nuclear spins with interaction strengths that are an order
of magnitude smaller than the electron spin dephasing rate. Our
results provide a route towards tomography with
single-nuclear-spin sensitivity and greatly extend the number of
available quantum bits for quantum information processing in
diamond.
\end{abstract}

\maketitle

Detecting the weak magnetic moment of a single nuclear spin
presents the ultimate limit of sensitivity in magnetic resonance
imaging \cite{Degen2009,Taylor2008c,Degen2008}. Furthermore,
nuclear spins may play a key role as qubits with long coherence
times in quantum information technologies \cite{Ladd2010}.
Addressing and controlling single nuclear spins is challenging
because the spins are generally embedded in a noisy environment,
such as a surrounding bath of nuclear spins.

The electron spin of a nitrogen-vacancy (NV) center is a powerful
probe of its local magnetic
environment~\cite{Maze2008,Taylor2008c,Balasubramanian2008,
Degen2008,DeLange2011,Zhao2011,Rondin2012,Kolkowitz2012,Hall2010}.
If a single or a few nuclear spins are located particularly close
to an NV center, the hyperfine interaction can well exceed the
electron spin dephasing rate 1/$T_2^*$ \cite{Childress2006}. Such
strongly coupled nuclear spins are readily distinguished from the
rest of the spin bath \cite{Smeltzer2011, Dreau2012} and can be
selectively addressed and controlled \cite{Dutt2007, Neumann2008,
Smeltzer2009, Neumann2010, Steiner2010, Fuchs2011, Jiang2009,
Robledo2011}. However, typically the nuclear spin of interest is
embedded in a bath of fluctuating nuclear spins. As a result, the
coupling of this single nuclear spin to the NV center is weak
compared to the rate of electron spin dephasing induced by the
spin bath. For both magnetometry and quantum information purposes
it would be greatly beneficial to be able to individually resolve
and address such weakly coupled nuclear spins.

\begin{figure}[b!]
    \includegraphics[scale=0.85]{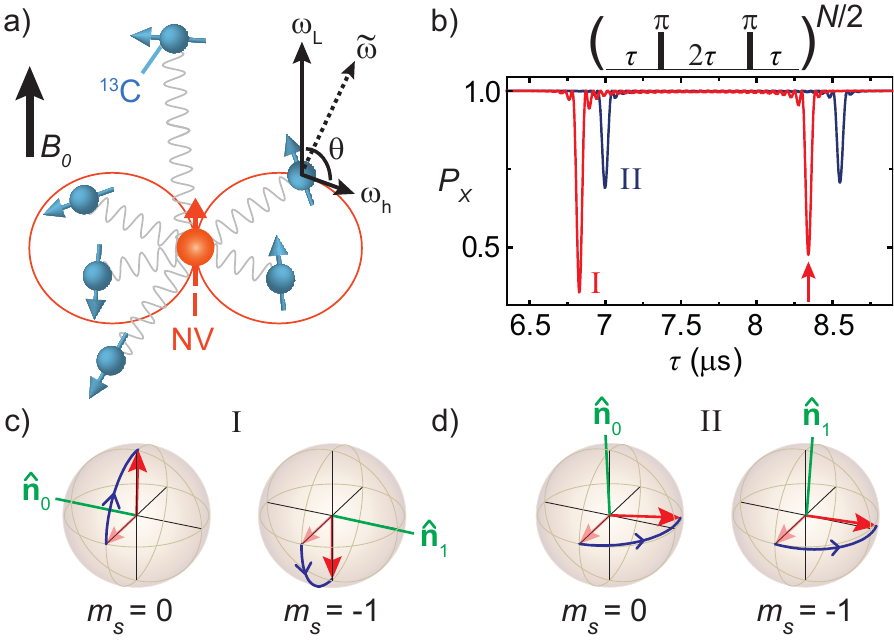}
    \caption{\label{Figure1} Concept of isolating and controlling weakly coupled spins.
    (a) Surrounding $^{13}$C nuclear spins precess about axes that depend on the NV electron spin state.
    For $m_s=0$, all $^{13}$C spins precess about $\boldsymbol{\omega}_L$ set by the applied magnetic field $\mathbf{B}_0$.
    For $m_s=-1$, each spin precesses about a distinct axes $\tilde{\boldsymbol{\omega}}$ due to the hyperfine interaction $\boldsymbol{\omega}_{h}$.
    (b) Calculated probability $P_x$ to preserve the initial electron spin state after a decoupling sequence with $N=32$,
    for two $^{13}$C spins with $\theta = \pi/4.5$, Nucleus I: $\omega_h
    = 2\pi\cdot40$ kHz, Nucleus II: $\omega_h = 2\pi\cdot20$ kHz, $B_0$=293~G.
    Each spin can be selectively addressed by tuning the inter-pulse delay $2\tau$
    into resonance with its dynamics. (c,d) Bloch spheres showing the nuclear spin dynamics for $\tau$ resonant with nucleus I (arrow).
    (c) For nucleus I, the net result is a rotation around anti-parallel
    axes ($\hat{\mathbf{n_0}}$ and $\hat{\mathbf{n_1}}$) for the two electron states, resulting in entanglement.
    (d) Nucleus II is decoupled: its rotation is independent of the electron state.}
\end{figure}

In this Letter, we isolate, characterize and selectively control
up to six weakly coupled $^{13}$C nuclear spins that are embedded
in the spin bath surrounding an NV center. The weak signal of a
specific nuclear spin is amplified by precisely tuning a
multi-pulse control sequence on the NV electron spin into
resonance with the electron-nuclear spin dynamics
\cite{VanderSar2012}. At the same time this sequence dynamically
decouples the electron spin from all other nuclear spins
\cite{Delange2010,Ryan2010,Naydenov2010}. With this technique, we
are able to resolve and coherently control nuclear spins with
couplings that are an order of magnitude smaller than the
dephasing rate of the NV center. Our results can enable tomography with single
nuclear spin sensitivity and have the potential to greatly extend
the number of solid-state spin qubits available for quantum
information processing.

Our method to isolate a weakly coupled nuclear spin from a
background of other nuclear spins is based on the distinct
conditional precession of each nuclear spin due to its particular
hyperfine interaction with the NV electron spin ($S=1$), Fig.\
1(a). For the electron in $m_s=0$, all nuclear spins precess with
the Larmor frequency $\omega_L$ around an axis parallel to the
applied magnetic field $\textbf{B}_0$. For $m_s=-1$, each nuclear
spin precesses around a distinct axis $\tilde{\boldsymbol{\omega}}
= \boldsymbol{\omega}_L + \boldsymbol{\omega}_h$. The hyperfine
interaction $\boldsymbol{\omega}_h$ depends on the position of
that particular nuclear spin relative to the NV center.

We can probe this conditional interaction by preparing the
electron spin in a superposition, $|x\rangle = (|m_s\! =\!
0\rangle + |m_s \! =\! -1\rangle)/\sqrt{2}$, and applying a
dynamical decoupling sequence consisting of $N$ sequential
$\pi$-pulses. Consider the basic decoupling unit on the electron
spin $\tau-\pi-2\tau-\pi-\tau$, in which $\tau$ is a free
evolution time (Fig.\ 1(b)). The net result of this unit is a
rotation of the nuclear spin by an angle $\phi$ around an axis
$\mathbf{\hat{n}}_i$ that depends on the initial state of the
electron spin: $\mathbf{\hat{n}}_0$ for initial state $m_s=0$ and
$\mathbf{\hat{n}}_1$ for initial state $m_s=-1$
\cite{VanderSar2012,SOM}.

If $\mathbf{\hat{n}}_0$ and $\mathbf{\hat{n}}_1$ are not parallel,
the resulting conditional rotation of the nuclear spin generally
entangles the electron and nuclear spins. As a result, for an
unpolarized nuclear spin state, the final electron spin state is a
statistical mixture of $|x\rangle$ and $|\!-x\rangle=(|m_s\!\!
=\!\! 0\rangle - |m_s \!\! =\!\! -1\rangle)/\sqrt{2}$. The
probability that the initial state $|x\rangle$ is preserved is
given by
    \begin{equation}\label{Px}
    P_x = (M+1)/2,
    \end{equation}
with for a single nuclear spin:
    \begin{equation}\label{Signal Modulation}
    M = 1 - (1 -
    \mathbf{\hat{n}}_0 \cdot \mathbf{\hat{n}}_1)\sin^2{\frac{N\phi}{2}}.
    \end{equation}
For multiple nuclear spins that do not mutually interact, $M$ is
given by the product of all the individual values $M_j$ for each
individual spin $j$. Analytical expressions for $\phi$ and for the
angle between $\mathbf{\hat{n}}_0$ and $\mathbf{\hat{n}}_1$ as a
function of the hyperfine interaction $\boldsymbol{\omega}_h$ and
the inter-pulse delay $\tau$ are given in the supplemental
material \cite{SOM}.

As an example, Fig.\ 1(b) shows calculated results for two
$^{13}$C spins with different hyperfine interactions. For most
values of $\tau$ the NV spin is effectively decoupled from both
nuclear spins and its initial state is conserved ($P_x \approx
1$). For specific values of $\tau$, the sequence is precisely
resonant for one of the $^{13}$C spins and a sharp dip in the
signal is observed. Figures 1(c, d) illustrate the evolution of
the nuclear spins at the resonance condition for nuclear spin I.
At this value of $\tau$, the net rotation axes
$\mathbf{\hat{n}}_0$ and $\mathbf{\hat{n}}_1$ for nuclear spin I
are approximately \emph{anti}-parallel and the resulting
conditional rotation entangles nuclear spin I with the electron
spin ($P_x \approx 1/2$). In contrast, at the same value of
$\tau$, $\mathbf{\hat{n}}_0$ and $\mathbf{\hat{n}}_1$ are nearly
parallel for nuclear spin II and the resulting unconditional
rotation leaves the electron spin unaffected. These resonances
appear periodically as a function of $\tau$.

More insight into the periodicity and depth of the resonances can
be gained by considering the case of large magnetic field,
$\omega_L \gg \omega_h$. In this case the positions of the
resonances are given by \cite{SOM}:
    \begin{equation}\label{Position}
    \tau_k = \frac{(2k-1)\pi}{2\omega_L + A},
    \end{equation}
where $k=1,2,3..$ is the order of the resonance, and $A$ is the
parallel component of the hyperfine interaction $A =
\omega_h\cos\theta$. Equation~\ref{Position} shows that the
position is a linear function of $k$. The amplitude of the
resonances is governed by the rotation angle $\phi$, which is of
order $B/\omega_L$, with $B=\omega_h\sin\theta$ the perpendicular
component of the hyperfine coupling. Although $\phi$ is small, the
total angle is amplified by the large number of pulses $N$,
enabling the detection with maximum contrast even of weakly
coupled spins. In this way a single nuclear spin can be isolated
from a bath of spins by a judicious choice of the inter-pulse
delay $2\tau$ and the number of pulses $N$.

    \begin{figure*}[tb]
    \includegraphics[scale=0.65]{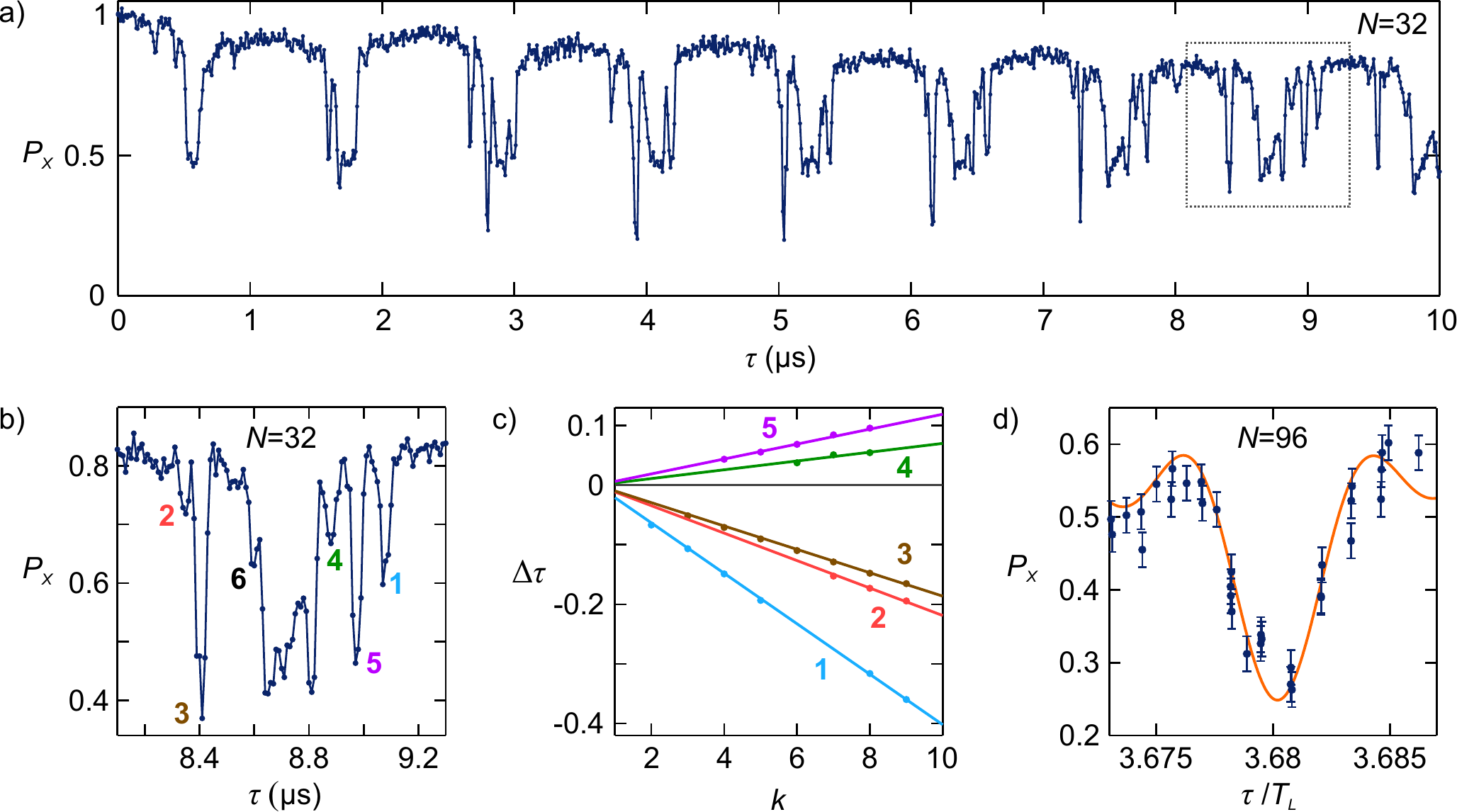}
    \caption{\label{Figure2}
    Resolving individual weakly coupled $^{13}$C nuclear spins. (a) $P_x$
    as function of $\tau$ for a decoupling sequence with $N=32$
    and a magnetic field $B_0=401$ G. The sharp resonances in the echo signal
    correspond to the coherent interaction with individual $^{13}$C atoms.
    (b) Magnification of the section marked in (a) indicating resonances associated with six nuclear spins.
    (c) Positions $\tau_k$ of resonances with order $k$ observed in (a) relative to the Larmor period $T_L = 2\pi/\omega_L$,
    $\Delta\tau = \tau_k/T_L - (2k-1)/4$. The five sets of
    equally-spaced
    resonances correspond to the spins numbered in (b). Lines are fits to Eq. \ref{Position}.
    (d) Close up for nuclear spin 6 ($\tau \approx 8.57...8.59\ \mu$s)
    with $N=96$. Line: fit based on equation \ref{Signal Modulation}.
    Errors are $\pm 1$ standard deviation (s.d.).}
    \end{figure*}

We experimentally demonstrate our method using an NV center in a
type IIa diamond with a natural abundance of $^{13}$C nuclear
spins ($1.1\%$). All experiments are performed at room temperature
with an applied magnetic field along the NV symmetry axis. The NV
electron spin is prepared in $m_s=0$ by illumination with a $532$
nm laser and read out through its spin-dependent fluorescence. The
experimental setup is described in detail in Ref.\ 23.

We choose an NV center that shows no nearby strongly coupled
$^{13}$C spins in the electron spin resonance (ESR) spectrum and
Ramsey measurements. The hyperfine coupling to the NV spin of all
individual $^{13}$C spins is thus weak compared to $1/T_2^*$: all
individual nuclear spins are hidden in the spin bath.

The experimental signal for a decoupling sequence with $32$
$\pi$-pulses is shown in Fig.\ 2(a). We observe sharp dips and
broader collapses in an approximately exponentially decaying signal
(see Fig.\ 2(b) for a magnification). The broader collapses
correspond to the overlapping signals of multiple nuclear spins in
the spin bath, whose product tends to yield $P_x \approx 0.5$. The
sharp dips are signatures of the resonances of individual $^{13}$C
spins. These appear primarily for large $\tau$ because the
separation between resonances of different spins increases
with the resonance order $k$ (see Eq.~\ref{Position}). We
exploit the linear dependence in Eq.~\ref{Position} to identify
five distinct $^{13}$C spins (Fig.~2(c)). The resonances assigned
to these spins are indicated in Fig.~2(b).

With a fit based on Eq.~\ref{Signal Modulation} we are able to
determine both the magnitude $\omega_h$ and the angle $\theta$ of
the hyperfine coupling from the experimentally observed resonances
in Fig.~2(b) for each of the five spins. These fits take the
overall signal decay due to relaxation to $m_s=+1$ and dephasing
of the electron state into account \cite{SOM}. Although nuclear
spin $6$ can not be clearly resolved from the spin bath with a
sequence of $32$ pulses (Fig.\ 2(b)), we can further increase the
sensitivity by applying more pulses. For $N=96$ the signal for
spin $6$ is well-isolated from the spin bath (Fig.\ 2(d)),
enabling the characterization of the hyperfine interaction.

The obtained values for the hyperfine interaction strength
$\omega_h$ and angle $\theta$ for the six nuclear spins are listed
in Table 1. These values should be compared to the minimal
coupling that can be resolved in an ESR measurement, which is
given by the ESR line width. We find that our method detects
hyperfine strengths as small as $\sim 20$ kHz, about an order of
magnitude smaller than the measured line width of $\sqrt{2}/(\pi
T_2^*) = 161(1)$ kHz. Furthermore, we resolve differences in
hyperfine strength down to $\sim 10$ kHz.

Assuming that the interaction is purely dipole-dipole, the values
in Table 1 correspond to distances to the NV center between $0.6$
and $1.2$ nm. The fact that we can distinguish multiple weakly
coupled spins beyond those that are coupled strongest to the NV
demonstrates that our method can be used to create tomographic
images of the spin environment at the single nuclear spin level.
\begin{table}
  \centering
   \begin{tabular}{c|c|c}
      Spin  & $\omega_h/2\pi$\ (kHz) & $\theta$\ (degrees) \\[1mm]
      \hline
      1 & 83.8(6) & 21(1) \\
      2 & 47(2) & 30(5)   \\
      3 & 55(2) & 54(2)   \\
      4 & 19(1) & 133(3)  \\
      5 & 33(1) & 132(1)  \\
      6 & 25.1(7) & 51(2) \\
   \end{tabular}
  \caption{Hyperfine coupling strength $\omega_h$ and angle
  $\theta$ for the six nuclear spins identified in Fig.~2. These
  values were obtained
  by fitting a single well-isolated resonance for each nuclear spin
  based on Eq.~\ref{Signal Modulation}.
  Uncertainties are 2 s.d.}\label{Table1}
\end{table}

    \begin{figure}[htb]
    \includegraphics[scale=0.65]{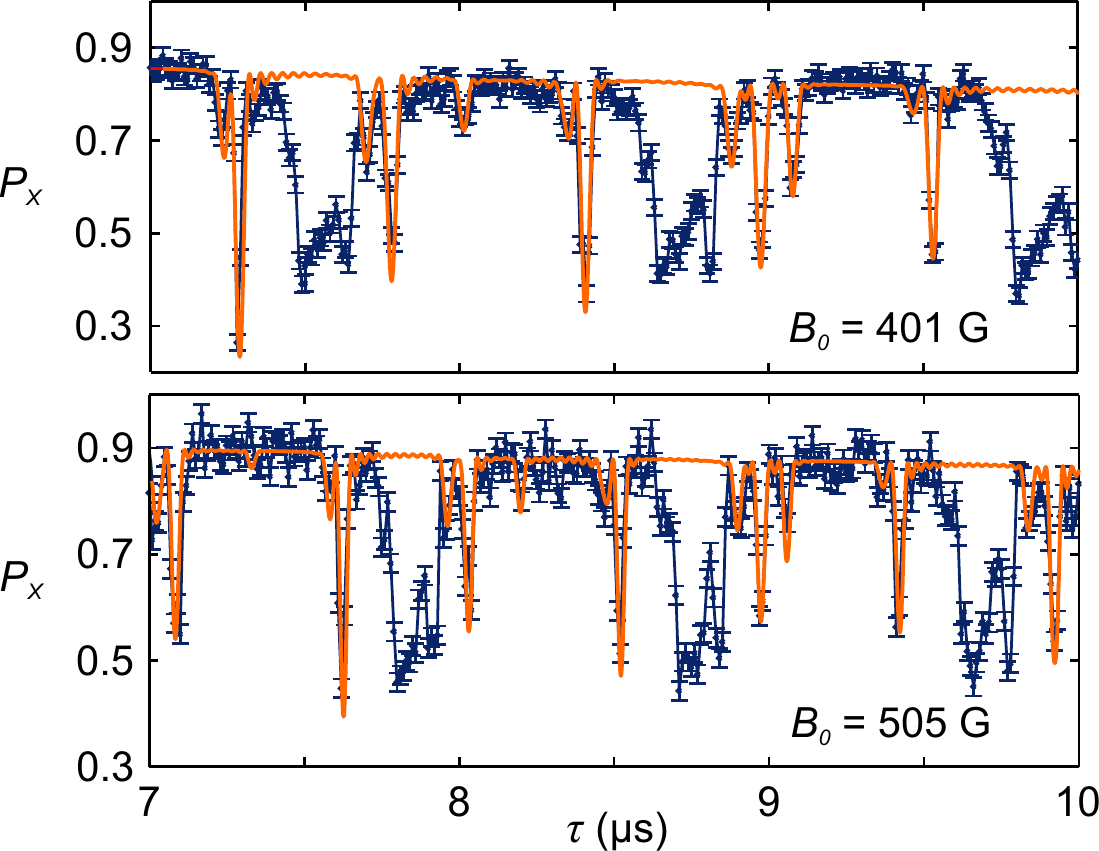}
    \caption{\label{Figure3}
    Comparison of the measured signal with the prediction based on the parameters in Table
    1 (orange line). We observe good agreement for the positions and amplitudes
    of multiple resonances for magnetic fields of both $B_0=401$ G
    and $B_0=505$ G. Error bars are $\pm 1$ s.d.}
    \end{figure}

We validate our approach by calculating the signal expected from
the values in Table 1, and comparing the result with independent
measurements over a broad range of free evolution times at two
different magnetic fields (Fig.~3). We find excellent agreement
for both the positions and amplitudes of the resonances,
confirming the accuracy of the theoretical model and the
determined parameters.

    \begin{figure}[htb]
    \includegraphics[scale=0.65]{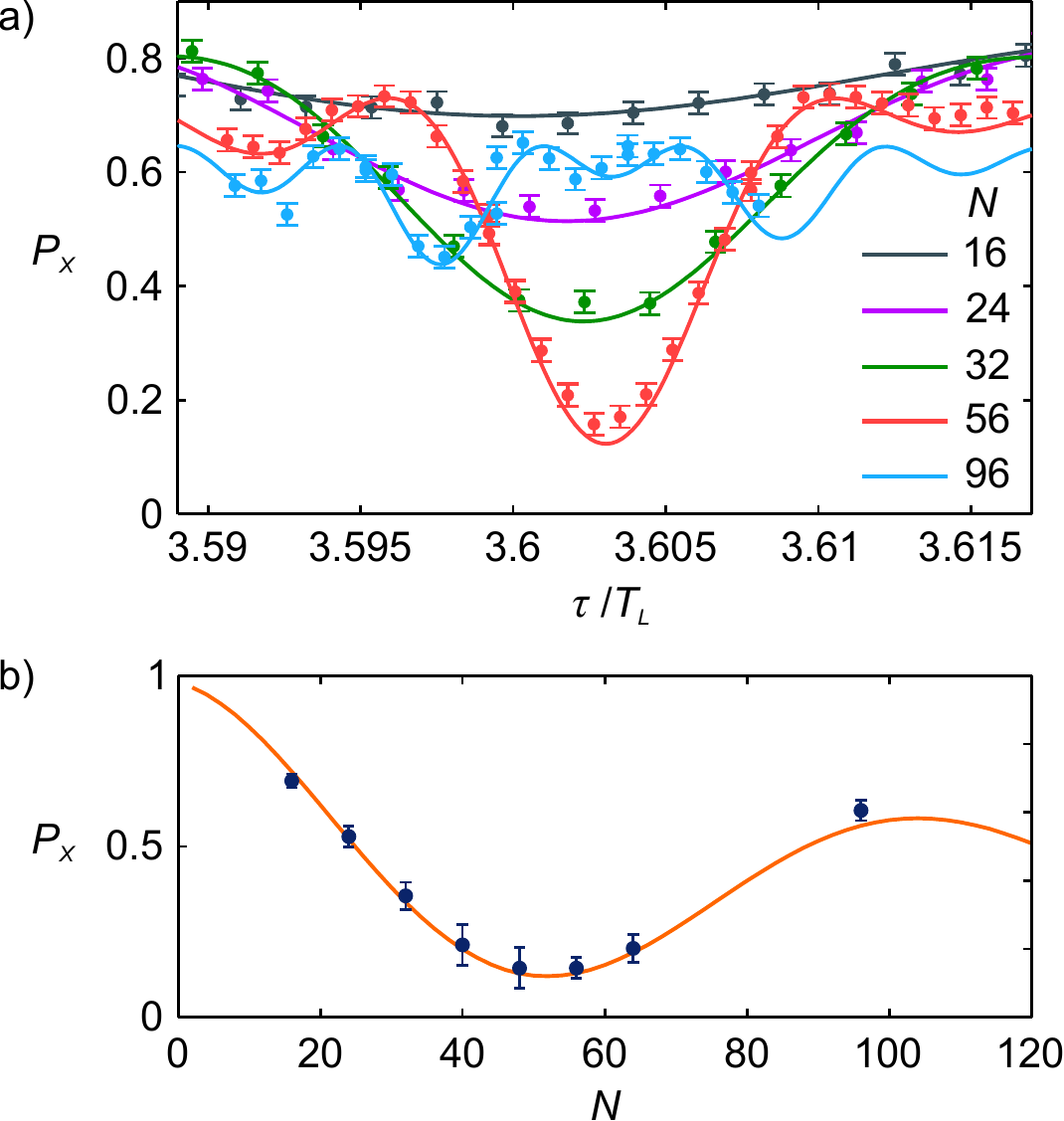}
    \caption{\label{Figure4}
    Coherent conditional rotations of weakly-coupled nuclear spin 3. (a) $P_x$ with different
    numbers of pulses $N$ for the resonance centered at $\tau\approx8.4\ \mu$s.
    Lines: calculations based on the values in Table 1. Error bars: $\pm 1$ s.d.
    (b) $P_x$ at resonance as function of $N$. The experimental
    values are obtained from Gaussian fits. A fit using the values in Table 1 (line) yields
    a decay constant for the oscillation of $1.8(3)$ ms (supplementary
    information). Uncertainties are $2$
    s.d.} \end{figure}

Finally, we demonstrate that we can coherently rotate a weakly
coupled nuclear spin over a desired angle by tuning the number of pulses $N$. Figure 4(a) plots the signal for a selected
resonance ($k=8$) of spin $3$ for different number of pulses $N$.
The depth of the resonance first increases with $N$ until the
maximum contrast is obtained for $N = \pi/\phi \approx 56$. For
more pulses the depth decreases again. In Fig. 4(b) we plot the signal at the center of the resonance as a function of $N$.

The oscillation observed in Fig.~4(b) demonstrates the coherent
conditional rotation of a single weakly coupled $^{13}$C spin. For
$N=28$ the signal reaches $\sim 0.5$. Here, the nuclear spin has
rotated over an angle $N\phi/2\approx \pi/2$, in a direction which
is conditioned by the electron spin state (similar to the case
illustrated in Fig.~1(c)). This sequence corresponds to a
maximally entangling operation, equivalent to the quantum
controlled-NOT gate up to single-qubit rotations. For $N=56$, the
nuclear spin has rotated over an angle $N\phi/2\approx\pi$. Here,
the two conditional rotations lead to the same final nuclear spin
state up to a 2$\pi$ phase difference. This phase difference
transfers to the electron spin, yielding the pure state
$|-x\rangle$ and signal $P_x\approx 0$.

Unconditional rotations of the nuclear spin can be implemented by
using different values for $\tau$ (see e.g. Fig.~1(d))
\cite{VanderSar2012}. A combination of conditional and
unconditional operations can be used to initialize the nuclear
spin by swapping its state with the electron~\cite{Dutt2007} or
for reading out the nuclear spin state in a single-shot by mapping
it onto the electron
spin~\cite{Jiang2009,Neumann2010,Robledo2011}. Our results thus
indicate the possibility of using weakly coupled nuclear spins as
fully controllable qubits.

The oscillation in Fig.~4(b) is damped on a timescale of a few ms.
This timescale is consistent with the longitudinal relaxation of
the electron spin at room temperature ($T_1$
process)~\cite{Jarmola2012}. At cryogenic temperatures this
relaxation time exceeds seconds \cite{Jarmola2012}, allowing for
the implementation of high-precision quantum gates on weakly
coupled $^{13}$C nuclei.

In conclusion, we have isolated, characterized and coherently
controlled individual weakly coupled nuclear spins embedded in a
spin bath. Because we address spins beyond the few nearest to the
NV center, our method can enable the tomography of ensembles of
spins in diamond and, potentially, in external samples
\cite{Cai2011}. In addition, the method enables coherent gates
between the electron spin and weakly-coupled nuclear spins. Our
results thus indicate a clear pathway for using weakly coupled
nuclear spins as qubits, thereby eliminating the need for strong
coupling and greatly extending the possible number of qubits
within a local register.

\textsl{Author's note -} While finalizing this manuscript we
became aware of two complementary studies that consider the
sensing of weakly-coupled nuclear spins in the low magnetic field
regime \cite{Kolkowitz2012a} and in isotopically purified diamond
\cite{Zhao2012}.

\textsl{Acknowledgements -} We thank S. Kolkowitz, M. D. Lukin, G. de Lange, H.
Fedder and J. Wrachtrup for discussions. This work is supported by
the Dutch Organization for Fundamental Research on Matter (FOM),
the Netherlands Organization for Scientific Research (NWO), the
DARPA QuEST program and the EU STREP program DIAMANT. THT
acknowledges support by a Marie Curie Intra European Fellowship.




%
%
%

\newpage
\
\newpage

\section{\textbf{Supplemental Material}}

\section{System dynamics during the decoupling sequence}
In this section we derive the equations used in the main text. We
consider the conditional dynamics of a nuclear spin that interacts
with the NV electron spin through the hyperfine coupling.

\subsection{Hamiltonian of the system}
With an appropriate rotation of the coordinate axes, the
Hamiltonian of the NV spin coupled to a single $^{13}$C spin is
given by
\begin{equation}
\hat{H}= A {\hat S_z}  \hat{I_z} + B \hat{S_z} \hat{I_x} +
\omega_L \hat{I_z} = |0\rangle\langle 0| \hat{H_0} +
|1\rangle\langle 1| \hat{H_1},
\end{equation}
where $\hat{S_i}$ ($\hat{I_i}$) are the Pauli matrices of the
electron (nuclear) spin, $\omega_L$ is the nuclear Larmor
frequency, and $A=\omega_h \cos \theta$ ($B=\omega_h \sin \theta$)
is the parallel (perpendicular) component of the hyperfine
coupling (see Fig. 1(a)). The nuclear spin evolves according to a
Hamiltonian which is conditioned by the electron spin state:
$\hat{H_0}$ if the electron is in $m_s=0$ (state $|0\rangle$), and
$\hat{H_1}$  if the electron is in $m_s=-1$ (state $|1\rangle$),
with
\begin{equation}
\hat{H_0}=\omega_L \hat{I_z}, \quad \text{and} \quad
\hat{H_1}=(A+\omega_L) \hat{I_z} + B \hat{I_x}.
\end{equation}

\subsection{Net result of the decoupling sequence}
As described in the main text, we probe the conditional
interaction by preparing the electron spin in $|x \rangle =
(|0\rangle + |1\rangle)/\sqrt{2}$, driving it by a decoupling
sequence with $N/2$ decoupling units of the form
$\tau-\pi-2\tau-\pi-\tau$ (Fig. 1(b)), and measuring the
$x$-component of the NV electron spin at the end of the sequence.

The probability that the initial state $|x \rangle$ is preserved
is given by $P_x=(M+1)/2$, with
\begin{equation}
M = {\rm Re}\ {\rm Tr}(\hat{U_0}\hat{U_1}^{\dag}),
\end{equation}
where $U_0$ and $U_1$ are the conditional evolution operators for
the nuclear spin corresponding to the initial states of the NV
spin $m_s=0$ and $m_s=-1$ respectively. These operators,
describing the evolution after $N/2$ decoupling units, have the
form $U_0=\hat{V_0}^{N/2}$ and $U_1=\hat{V_1}^{N/2}$, where
$\hat{V_0}$ and $\hat{V_1}$ are the conditional evolution
operators for a single decoupling unit ($N=2$):
\begin{eqnarray}
\label{eq:V01}
\hat{V_0}&=&\exp{[-i\hat{H_0} \tau]} \exp{[-i \hat{H_1} 2\tau]} \exp{[-i\hat{H_0} \tau]}\\
\hat{V_1}&=&\exp{[-i\hat{H_1} \tau]} \exp{[-i \hat{H_0} 2\tau]}
\exp{[-i\hat{H_1} \tau].}
\end{eqnarray}
Since any unitary evolution of a single qubit is a rotation, the
conditional operators $\hat{V_0}$ and $\hat{V_1}$ can be
represented as
\begin{eqnarray}
\hat{V_0}&=&\exp{[-i\phi (\mathbf{\hat I} \cdot \mathbf{\hat{n}}_0)]}\\
\hat{V_1}&=&\exp{[-i\phi (\mathbf{\hat I} \cdot
\mathbf{\hat{n}}_1)],} \label{eq:V02}
\end{eqnarray}
which illustrates that the net evolution of the nuclear spin after
a single decoupling unit is a rotation by an angle $\phi$ around
an axis $\mathbf{\hat{n}}_i$ that depends on the initial state of
the electron spin: $\mathbf{\hat{n}}_0$ for initial state $m_s=0$
and $\mathbf{\hat{n}}_1$ for initial state $m_s=-1$. Note that the
rotation angle $\phi$ is independent of the electron spin input
state because $\cos\phi = {\rm Tr} \hat{V_0} = {\rm Tr}
\hat{V_1}$.

With these expressions, we obtain Eq.~2 of the main text:
\begin{equation}
\label{Signal Modulation_SUP} M = 1 - (1 - \mathbf{\hat{n}}_0
\cdot \mathbf{\hat{n}}_1)\sin^2{\frac{N\phi}{2}}.
\end{equation}
Thus, we only need to determine the net rotation angle $\phi$ and
the inner product $\mathbf{\hat{n}}_0 \cdot \mathbf{\hat{n}}_1$ to
completely characterize the signal.
\subsection{Nuclear spin rotation axes and angle}
Using eqs.~\ref{eq:V01}-~\ref{eq:V02}, we obtain expressions for
the rotation angle $\phi$ and the inner product
$\mathbf{\hat{n}}_0 \cdot \mathbf{\hat{n}}_1$:
\begin{eqnarray}
\cos{\phi} &=& \cos{\alpha}\cos{\beta} - m_z\sin{\alpha}\sin{\beta}\\
1-\mathbf{\hat{n}}_0 \cdot \mathbf{\hat{n}}_1 &=& m_x^2\
\frac{(1-\cos{\alpha})(1-\cos{\beta})}
{1+\cos{\alpha}\cos{\beta}-m_z\sin{\alpha}\sin{\beta}},
\end{eqnarray}
where $m_z=(A+\omega_L)/\tilde{\omega}$, $m_x=B/\tilde{\omega}$,
and $\tilde{\omega}=\sqrt{(A+\omega_L)^2 +B^2}$. For simplicity,
we introduced the angles $\alpha=\tilde{\omega} \tau$ and
$\beta=\omega_L \tau$.

\subsection{High magnetic field: $\omega_L \gg \omega_h$}
The advantage of performing the experiments at high magnetic
field, so that $\omega_L \gg \omega_h$, is that the signal $M$ has
a clear form: for most values of $\tau$, the depth of the signal
modulation $(1-\mathbf{\hat{n}}_0 \cdot \mathbf{\hat{n}}_1)$ is
small, of order of $m_x^2\sim B^2/\omega_L^2$. Here, the axes
$\mathbf{\hat{n}}_0$ and $\mathbf{\hat{n}}_1$  are practically
parallel and the nuclear spin undergoes an unconditional rotation
(Fig 1(d)). The modulation is only strong when the decoupling
sequence is precisely resonant with the nuclear spin dynamics. At
resonance, the axes $\mathbf{\hat{n}}_0$ and $\mathbf{\hat{n}}_1$
are anti-parallel, i.e. $\mathbf{\hat{n}}_0 \cdot
\mathbf{\hat{n}}_1 = -1$, leading to a conditional rotation of the
nuclear spin (Fig. 1(c)). The condition for the resonances is
\begin{equation}
\label{eq:cond}
\tan{\frac{\alpha}{2}}\,\tan{\frac{\beta}{2}}=\frac{1}{m_z}.
\end{equation}

For the high magnetic field case where $m_x\ll 1$, in first order
in this small quantity, the solution for Eq.~\ref{eq:cond} is
$\alpha=-(\beta+\pi) + 2k\pi$ with any integer $k$, so that
$\tan{(\alpha/2)}=1/\tan{(\beta/2)}$, and the corrections are of
the second order in $m_x$. In this case the inter-pulse distance
$2\tau$ corresponding to the resonance condition satisfies
\begin{equation}
\label{eq:condition} \tau=\tau_k \approx
\frac{(2k+1)\pi}{2\omega_L + A},
\end{equation}
which is equivalent to the Eq.~3 of the main text.

We now provide more insight into the shape and amplitude of the
resonances. To understand what happens when $\tau$ is varied from
the resonant value $\tau_k$, we assume that $\tau=\tau_k
(1+\Delta)$ with small $\Delta$. Then, the inner product of the
rotation axes becomes
\begin{equation}
1-(\mathbf{\hat{n}}_0 \cdot \mathbf{\hat{n}}_1) = 2 / (1 +
\delta_k^2/m_x^2),
\end{equation}
where $\delta_k=(2k+1)\pi\Delta$, and we assume as above that
$A,B\ll \omega_L$. Thus, this modulation depth as a function of
$\Delta$ is a Lorentzian with a small width of $m_x/[(2k+1)\pi]$
(this justifies our assumption about $\Delta$ being small, not
exceeding $m_x\ll 1$ too much). The final result as a function of
$\delta_k$ acquires the form
\begin{equation}
M=1 - \frac{2}{1 + \delta_k^2/m_x^2}\
\sin^2{\left[\frac{N}{2}\sqrt{m_x^2+\delta_k^2}\right]}.
\end{equation}
This expression demonstrates that even when $m_x\ll 1$ the
resonance can be visible. When the argument of the sine function
above is small, the amplitude of the resonance is of order $m_x^2
N^2$, and is therefore amplified by a factor of $N^2$. By
increasing the number of pulses, we increase the sensitivity of
the detection of the nuclear spins, i.e.\ we can address more and
more nuclei, which are coupled weaker and weaker to the NV spin.

\section{Electron spin relaxation and dephasing}

For the fits in the main text we take the overall decay of the
signal due to the relaxation and dephasing of the electron spin
into account. The equation relating the obtained signal $P_x$ to
the modulation $M$ due to the interaction with the nuclear spins
becomes:
\begin{equation}\label{relaxation}
P_x = 1/2e^{-2N\tau/T_\alpha}M + 1/3 + 1/6e^{-2N\tau/T_\beta},
\end{equation}
in which $T_\alpha$ and $T_\beta$ are two phenomenological time
constants. Note that at $\tau=0$ the signal modulation is maximum
and equation \ref{relaxation} is equivalent to the ideal case
given in the text.

$T_\beta$ describes the $T_1$ relaxation of the electron spin into
the $m_s=+1$ sublevel. This relaxation affects the center value
that the modulation $M$ acts on. We measure this time constant by
taking the average value of the outcomes of measurements along $x$
and $-x$ ($|-x \rangle = (|0\rangle - |1\rangle)/\sqrt{2}$), i.e.
effectively we measure for $M=0$. We obtain $T_\beta = 2.2$ ms.

$T_\alpha$ accounts for dephasing of the electron spin and
includes not-decoupled dephasing, microwave pulse errors and
decoherence due to $T_1$ relaxation. These effects reduce the
total modulation amplitude. This reduction is only approximately
exponential, but this is a good approximation over the small
$\tau$ ranges that are used for the fits. The values obtained from
fits vary between $1.3$ and $2.2$ ms for different measurements
depending on the number of pulses, the fidelity of the
$\pi$-pulses, and the applied magnetic field.

\end{document}